\documentstyle[12pt]{article}

\input epsf.sty

\newcommand{\g}{{\sl g}}

\newcommand{\D}{{\cal D}}
\newcommand{\G}{{\cal G}}

\newcommand{\K}{{\cal K}}
\newcommand{\N}{{\cal N}}
\newcommand{\Q}{{\cal Q}}
\newcommand{\V}{{\cal V}}
\newcommand{\U}{{\cal U}}

\newcommand{\X}{{\cal X}}

\def\1{\hbox{{1}\kern-.25em\hbox{l}}}

\begin{document}

\begin{titlepage}

{\hfill \parbox{20mm}{TPR-32-98}}

\vspace{3mm}

\centerline{\large \bf Implications of $\ {\cal N} = 1$
                       supersymmetry }
\centerline{\large \bf for QCD conformal operators. }

\vspace{15mm}

\centerline{\bf A.V. Belitsky\footnote{Alexander von Humboldt Fellow.},
            D. M\"uller, A. Sch\"afer}

\vspace{15mm}

\centerline{\it Institut f\"ur Theoretische Physik, Universit\"at
                Regensburg}
\centerline{\it D-93040 Regensburg, Germany}

\vspace{20mm}

\centerline{\bf Abstract}

\hspace{0.5cm}

We prove a set of identities for the anomalous dimensions of the quark and
gluon conformal operators in the flavour singlet channel in QCD. These
relations arise from the graded commutator algebra of the $\ {\cal N} =
1$ superconformal group. We evaluate the rotation matrices for the 
quantities under study from the conventional dimensional regularization 
to the supersymmetry preserving regularization scheme. Using them we verify 
the equalities in two-loop approximation employing the results for the NLO 
anomalous dimensions of the conformal operators in the minimal subtraction 
scheme derived earlier.

\vspace{5cm}


\vspace{0.5cm}


\end{titlepage}


\noindent {\it 1. Introduction.} In spite of a number of attractive
theoretical features of supersymmetric gauge theories \cite{Son85}
in view of unification of the fundamental forces of Nature, the
manifestation of their predictions was not observed experimentally so
far. Nevertheless, they provide an excellent technical playground for 
the exploration of new physical concepts. It is remarkable that due to
the high symmetry of the underlying Lagrangian the theory enjoys the
property of reducibility of independent parameters such as gauge couplings 
and field renormalization constants\footnote{The latter holds true 
provided a supersymmetry preserving quantization and regularization 
procedures are used to handle the underlying Lagrangian.}. Since the 
structure of the corresponding Lagrangians resembles that of ordinary 
gauge theories with an adjusted fermion sector it can serve as a 
technical tool for deriving relationships between observables. Using 
heuristic arguments of this kind several relations have been derived 
in Ref.\ \cite{BFKL85} between the anomalous dimensions of the local 
quark and gluon operators without total derivatives which appear in the 
description of the deep inelastic scattering via the operator product 
expansion. One of them is the empirically established Dokshitzer 
relation \cite{Dok77}.

In this paper we address the issue of constraints on the anomalous
dimensions of the conformal operators from the point of view of graded 
commutator algebra of the superconformal group and Ward identities for 
the Green functions with composite operator insertion. Presently we 
consider restricted $\Q$-supersymmetry transformations which provide 
relations for the scale anomalies of composite operators with total 
derivatives, i.e.\ their anomalous dimensions. We make use of the 
subalgebra of the full superconformal algebra \cite{WesZumB70,Fer74} 
which includes scale transformation and consists of the relations, apart 
from the usual ones,
\begin{equation}
\label{GradedAlgebra}
[\Q,{\cal P}_\mu]_- = 0 , \quad
[\Q,{\cal M}_{\mu\nu}]_- = \frac{1}{2} \sigma_{\mu\nu} \Q , \quad
[\Q, \bar\Q]_+ = 2 \gamma_\mu {\cal P}_\mu , \quad
[\Q,\D]_- = \frac{i}{2} \Q .
\end{equation}
In the infinitesimal form their action on the field operator $\phi$ is
defined as $\delta^G \phi \equiv i [\phi, \G]_-$ with $\G = {\cal P}_\mu,
{\cal M}_{\mu\nu}, \D$ for even generators and $\G = \bar\zeta \Q$
for odd $\Q$ with $\zeta$ being a Majorana Grassman valued spinor.
One can equally include special supertransformations \cite{WesZumB70,Fer74}
which would allow to derive relations for the special conformal anomalies
\cite{BelMul98QCD} by means of the anomalous superconformal Ward identities.


\noindent {\it 2. ${\cal N} = 1$ QCD.} The Lagrangian for the $\N = 1$
super-Yang-Mills theory \cite{SUSY-YM} in the Wess-Zumino gauge
\cite{WesZumB78} takes the form which resembles ordinary one-flavour 
QCD with Majorana fermions, $\psi = C \bar\psi^{\rm T}$, in the adjoint 
representation of the $SU(N_c)$ group
\begin{equation}
\label{ClassicLagr}
{\cal L}_{\rm cl} = - \frac{1}{4} \left( G^a_{\mu\nu} \right)^2
+ \frac{i}{2} \bar\psi^a \not\!\!\D^{ab} \psi^b
+ \frac{1}{2} \left( D^a \right)^2 .
\end{equation}
The Wess-Zumino gauge breaks linear $\N = 1$ supersymmetry but the
remaining short supermultiplet $\left(B^a_\mu, \psi^a, D^a\right)$
respects the restricted non-linear supersymmetric transformation laws
\begin{equation}
\label{NonLinSUSY}
\delta^Q \psi^a = \frac{i}{2} G^a_{\mu\nu} \sigma_{\mu\nu} \zeta
- i D^a \gamma_5 \zeta, \quad
\delta^Q B^a_\mu = - i \bar\zeta \gamma_\mu \psi^a, \quad
\delta^Q D^a = \bar\zeta \not\!\!\D^{ab}\gamma_5 \psi^b .
\end{equation}
The action transforms w.r.t.\ these transformations as $\delta^Q S = -
\int d^4 x \left\{ \partial_\rho \bar\zeta Q_\rho - \sum_\phi \left(
\delta S / \delta \phi \right) \delta^Q \phi \right\}$ with the 
non-anomalous (on quantum level) \cite{Cur77} supersymmetry current 
$Q_\rho = \frac{1}{2} G^a_{\mu\nu} \sigma_{\mu\nu} \gamma_\rho \psi^a$
\cite{Zum74}. It is conserved on the mass shell: $\partial_\rho
\bar\zeta Q_\rho = \sum_\phi \left( \delta S / \delta \phi \right)
\delta^Q \phi$ and, thus, $\delta^Q S = 0$.

Eqs.\ (\ref{NonLinSUSY}) form, however, a modified algebra which
contains apart from the usual terms on the r.h.s.\ of (\ref{GradedAlgebra})
also a gauge transformation $\delta^{\rm gauge}$ \cite{WitFre75},
i.e.\ $[\delta_1^Q, \delta_2^Q]_- = - 2 i a_\mu \delta_\mu^P +
\delta^{\rm gauge}$, with field dependent gauge parameter,
$2 i \bar\zeta_1 \!\not\!\!B^a \zeta_2$, and translation
vector $a_\mu = \bar\zeta_1 \gamma_\mu \zeta_2$. Note, however,
that they define the ordinary SUSY operator algebra (\ref{GradedAlgebra})
in the space spanned on gauge invariant objects.

In our consequent discussion we will be most interested in the commutator
algebra restricted to the action on the ``good" light-cone components
\cite{KogSop70} of the field operators introduced as follows\footnote{The
$+$ and $-$ components of any vector are obtained by contraction with the
two light-like vectors $n$ and $n^*$, such that $n^2 = n^{*2} = 0$ and
$nn^* = 1$.} with projectors ${\mit \Pi}_{\pm} = \frac{1}{2} \gamma_{\mp}
\gamma_{\pm}$ for fermion fields: the $\pm$-components of which are defined 
via $\psi^a_{\pm} = {\mit \Pi}_\pm \psi^a$; and with the two-dimensional 
metric tensor $g^\perp_{\mu\nu} = g_{\mu\nu} - n_\mu n^\ast_\nu 
- n_\nu n^\ast_\mu$, for bosons: $B^{a \perp}_\mu = g^\perp_{\mu\nu} 
B^a_\nu$. We also impose a restriction on the constant Majorana fermion 
$\zeta_+ \equiv {\mit \Pi}_+ \zeta = 0$, so that $\delta^Q B^a_+ = 0$. Then 
a peculiar feature of the supersymmetry transformations (\ref{NonLinSUSY}) 
becomes manifest, namely, that they do not involve the auxiliary field, 
$D^a$, provided we restrict ourselves to the ``good" components --- the 
ones which enter the quasi-partonic conformal operators defined below 
in Eq.\ (\ref{treeCO}).
Explicitly \cite{LCsusyN4,LCsusyN1}
\begin{equation}
\label{LCLinSUSY}
\delta^Q \psi^a_+
= - G^{a \perp}_{+ \mu} \gamma_- \gamma^\perp_\mu \zeta , \quad
\delta^Q B^{a \perp}_\mu
= - i \bar\zeta \gamma^\perp_\mu \psi^a_+ .
\end{equation}
Moreover, they form a representation of the unaltered supersymmetry
algebra (\ref{GradedAlgebra}).


\noindent {\it 3. Supermultiplet of conformal operators.}
To derive the relations for the anomalous dimensions we have first
to find an irreducible representations of the superalgebra in the
basis of the conformal operators \cite{BelMul98QCD}:
\begin{equation}
\label{treeCO}
\left\{\!\!\!
\begin{array}{c}
{^Q\!{\cal O}^V} \\
{^Q\!{\cal O}^A}
\end{array}
\!\!\!\right\}_{jl}
\!= \frac{1}{2}
\bar\psi^a_+ (i \partial_+)^l\!
\left\{\!\!\!
\begin{array}{c}
\gamma_+ \\
\gamma_+ \gamma_5
\end{array}
\!\!\!\right\}
\!C^{3/2}_j\!
\left( \frac{\stackrel{\leftrightarrow}{\D}_+}{\partial_+} \right)
\!\psi^a_+ , \
\left\{\!\!\!
\begin{array}{c}
{^G\!{\cal O}^V} \\
{^G\!{\cal O}^A}
\end{array}
\!\!\!\right\}_{jl}
\!=
G^{a \perp}_{+ \mu} (i \partial_+)^{l-1}\!
\left\{\!\!\!
\begin{array}{c}
g_{\mu\nu} \\
i\epsilon_{\mu\nu-+}
\end{array}
\!\!\!\right\}
\!C^{5/2}_{j - 1}\!
\left(
\frac{\stackrel{\leftrightarrow}{\D}_+}{\partial_+}
\right)
\!G^{a \perp}_{\nu +},
\end{equation}
where $\partial \!= \stackrel{\rightarrow}{\partial}
\!\!+\!\! \stackrel{\leftarrow}{\partial}$
and  $\stackrel{\leftrightarrow}{\D}
= \stackrel{\rightarrow}{\D} - \stackrel{\leftarrow}{\D}$,
and the factor $\frac{1}{2}$ in front of the fermion operator serves
to avoid double counting of the same components of the spinors due to
their Majorana nature.

According to the appendix one can introduce the following combinations
of the conformal operators
\begin{equation}
\label{SUSYmultiplet}
\left\{\!\!\!
\begin{array}{c}
{\cal S}^1
 \\
{\cal P}^1
\end{array}
\!\!\!\right\}_{jl}
\equiv
\frac{6}{j}
{^G\!{\cal O}}^{\mit\Gamma}_{jl}
+
{^Q\!{\cal O}}^{\mit\Gamma}_{jl}, \qquad
\left\{\!\!\!
\begin{array}{c}
{\cal S}^2
 \\
{\cal P}^2
\end{array}
\!\!\!\right\}_{jl}
\equiv
\frac{6}{j + 1}
{^G\!{\cal O}}^{\mit\Gamma}_{jl}
-
\frac{j + 3}{j + 1}
{^Q\!{\cal O}}^{\mit\Gamma}_{jl},
\end{equation}
with ${\mit\Gamma} = V(A)$ for the upper (lower) entry on the l.h.s.\ of 
these equations. By construction they transform covariantly w.r.t.\
supertransformations and together with
\begin{equation}
\left\{\!\!\!
\begin{array}{c}
\V \\
\U
\end{array}
\!\!\!\right\}_{jl}
\equiv
\frac{(j + 2)(j + 3)}{(j + 1)}
G^{a \perp}_{+ \mu} (i \partial_+)^l
P^{(2,1)}_j \left(
\frac{\stackrel{\leftrightarrow}{\D}_+}{\partial_+}
\right)
\gamma^\perp_\mu
\left\{\!\!\!
\begin{array}{c}
1 \\
\gamma_5
\end{array}
\!\!\!\right\}
\psi^a_+ ,
\end{equation}
form an irreducible representation with the transformation laws:
\begin{eqnarray}
\label{SUSYtransfS}
&&\delta^Q\, {\cal S}^1_{jl} = \frac{1}{2} [1 - (-1)^j]\
\bar\zeta \V_{j - 1 l}, \qquad
\delta^Q\, {\cal S}^2_{jl} = \frac{1}{2} [1 - (-1)^j]\
\bar\zeta \V_{jl},\\
\label{SUSYtransfP}
&&\delta^Q\, {\cal P}^1_{jl} = \frac{1}{2} [1 + (-1)^j]\
\bar\zeta \U_{j - 1 l}, \qquad
\delta^Q\, {\cal P}^2_{jl} = \frac{1}{2} [1 + (-1)^j]\
\bar\zeta \U_{jl},\\
\label{SUSYtransfV}
&&\delta^Q \V_{j - 1l - 1}
= - \gamma_-\zeta
\left\{
{\cal S}^1_{jl} + {\cal S}^2_{j - 1 l}
\right\}
- \gamma_-\gamma_5\zeta
\left\{
{\cal P}^1_{jl} + {\cal P}^2_{j - 1 l}
\right\} .
\end{eqnarray}
The transformation law for the operator $\U_{jl}$ follows from
the observation $\U_{jl} = - \gamma_5 \V_{jl}$ and the Eq.\
(\ref{SUSYtransfV}).


\noindent {\it 4. Ward identities and commutator constraints.} To
proceed with our derivation of the Ward identities we have to fix the
remaining ordinary gauge degrees of freedom of the classical Lagrangian
(\ref{ClassicLagr}), i.e.\ to add a gauge fixing term together with an
associated ghost piece. Provided we would work in the superfield formalism
this can be done in a way which preserves linear supersymmetry with SUSY
Fermi-Feynman gauge. However, then we should proceed with the complete
gauge supermultiplet at an expense of a number of auxiliary fields on top
of the dynamical $\psi$ and $B$. To achieve this goal in the Wess-Zumino
supergauge without explicit breaking of the supersymmetry on the Lagrangian
level one is forced to use the light-cone gauge\footnote{Covariant gauges
inevitably break SUSY since the necessary condition for keeping
symmetry of the Lagrangian intact is that the algebra of the gauge
and rigid symmetries has the form of a semi-direct product \cite{Hol88}.
However, this breaking is only due to the BRST exact operator
$\delta^{\rm BRST}\delta^Q (\bar\omega^a \partial_\mu B^a_\mu)$ (since
$[\delta^{\rm BRST} , \delta^Q]_- = 0$ \cite{WitFre75,Fuj98}) and
will not affect physical quantities. We shall address this question
within the present context elsewhere. Next, the axial gauge
breaks the supersymmetry of the Lagrangian but leads to supersymmetric
counterterms for the gauge multiplet \cite{AntFlo83,CapJonPac86}.
However, the latter property is violated as soon as matter superfields
are taken into account \cite{CapJonPac86}. In spite of the fact that
we are considering only gauge multiplet we prefer to deal with
supersymmetry preserving gauge fixing on the Lagrangian level.}
\cite{LCsusyN4,KouRos83,LCsusyN1} so that
${\cal L} = {\cal L}_{\rm cl} + {\cal L}_{\rm gf} + {\cal L}_{\rm gh}$,
where ${\cal L}_{\rm gf} = - \frac{1}{2\xi}\left( B^a_+ \right)^2$,
${\cal L}_{\rm gh} = \bar\omega^a \D^{ab}_+ \omega^b$ \cite{Bas91}.
Ghosts decouple from all Green functions since $n_\mu \langle B^a_\mu
(x_1) B^b_\nu (x_2) \rangle |_{\xi \to 0}= 0$. However, because the 
Lorentz invariance is violated by fixing a particular direction
in the Minkowski space with the vector $n_\mu$, different (i.e.\ ``good"
and ``bad") components of the field operators renormalize with
different renormalization constants and, moreover, non-local counterterms
are required to cure all divergencies of the classical action, i.e.\
\cite{Bas91}
\begin{equation}
B^a_\mu \to Z_3^{1/2}
\left( B^a_\mu - (1 - \widetilde Z_3^{-1}) n_\mu {\mit\Omega}^a \right) ,\
\psi \to Z_2^{1/2}
\left( {\mit\Pi}_+ + \widetilde Z_2 {\mit\Pi}_- \right) \psi ,\
\g \to Z_\g \g ,\
\xi \to Z_3 \xi ,
\end{equation}
where ${\mit\Omega}^a = \left( {\cal D}^{-1}_+ G_{+-} \right)^a$ is
the only non-local structure involved. This instance complicates
the derivation and the use of the conformal Ward identities. However, 
the fact that only ``good" components are relevant suggests to 
integrate out the ``bad" ones in favour of the ``good" ones 
\cite{LCsusyN4,LCsusyN1,Bas96} in the path integral for the Green 
functions with operator insertions, ${\cal O}$,
\begin{equation}
\label{LCGreen}
\langle {\cal O} \X
\rangle = \lim_{\xi \to 0} \left( \int D\phi\ e^{iS} \right)^{-1}
\int D\phi\ {\cal O} \X e^{iS} ,
\end{equation}
where $S = \int d^d x {\cal L}$ and $\X$ is a monomial of the ``good"
components of the field operators, i.e.\ $\X = \prod_i \bar\psi_+ (x_i)
\prod_j \psi_+ (x_j) \prod_k B^\perp (x_k)$. In spite of the fact that
this procedure results in the non-local form of the action, the Lagrangian
manifests explicit supersymmetry w.r.t.\ renormalized transformations
(\ref{LCLinSUSY}) since from the Slavnov-Taylor identities it follows that
$Z_3 = Z_2 \equiv Z_\phi$ \cite{LCsusyN1}. To maintain
this property in perturbation theory it is necessary to deal with UV
divergencies in a supersymmetric way, i.e.\ by regularization via, e.g.\
Siegel's dimensional reduction\footnote{This is not a fully self-consistent
regularization method \cite{Sie80} but at least its reliability as a scheme
which respects supersymmetry has been checked on two-loop level
\cite{CapJonNie80}.} (DRED) \cite{Sie79,CapJonNie80,Sie80}. Provided
we have done this, the light-cone gauge preserves SUSY (\ref{LCLinSUSY})
on the quantum level as well and the transformations (\ref{LCLinSUSY})
remain unaffected even for the renormalized fields. This was checked in
one-loop approximation for the principal value (PV) \cite{KouRos83} and
the Mandelstam-Leibbrandt (ML) \cite{LCsusyN1} prescription on auxiliary
$\frac{1}{k_+}$-pole in the gluon propagator. In the latter case
$Z_\g = Z_3^{-1/2}$, while in the former $Z_{2,3}$-factors are
$\epsilon$-dependent due to the fact that the PV prescription violates
power counting. This can be traced back to the breaking of rescaling
invariance of the gluon density matrix.

It is obvious, that in the course of renormalization the operators
${\cal S}^i$ (${\cal P}^i$) mix with each other. We introduce the
renormalized operators according to
\begin{equation}
\label{renCO-OP}
[ \mbox{\boldmath${\cal O}$}_{jl} ]
= \sum_{k = 0}^{j}
\left\{\mbox{\boldmath$Z$}_{\cal O}\right\}_{jk}
\mbox{\boldmath$Z$}_\phi
\mbox{\boldmath${\cal O}$}^{(0)}_{kl},
\quad\mbox{with}\quad
\mbox{\boldmath$Z$}_{\cal O} =
\left({
{^{11}\!Z}_{\cal O}\ {^{12}\!Z}_{\cal O}\atop
{^{21}\!Z}_{\cal O}\ {^{22}\!Z}_{\cal O}
}\right), \quad
\mbox{\boldmath$Z$}_\phi =
\left({
Z_\phi^{-1} \ \ \ 0 \atop
0 \ \ \ Z_\phi^{-1}
}\right) ,
\end{equation}
where $\mbox{\boldmath${\cal O}$}_{jl}$ stands for a two-dimensional vector
$\mbox{\boldmath${\cal S}$} = \left( { {\cal S}^1 \atop {\cal S}^2 }
\right)$, and similar for ${\cal P}^i$. The superscript $(0)$ means that
the operator is written in terms of bare fields. And the anomalous
dimensions are defined via the renormalization group equations
\begin{eqnarray}
&&\mu \frac{d}{d\mu} [\mbox{\boldmath${\cal O}$}_{jl}]
= - \sum_{k = 0}^{j} \mbox{\boldmath${\gamma}$}^{\cal O}_{jk}
[\mbox{\boldmath${\cal O}$}_{kl}],
\quad\mbox{with}\quad
\mbox{\boldmath$\gamma$}^{\cal O} =
\left({
{^{11}\!\gamma}^{\cal O}\ {^{12}\!\gamma}^{\cal O}\atop
{^{21}\!\gamma}^{\cal O}\ {^{22}\!\gamma}^{\cal O}
}\right)
\quad\mbox{for}\quad
\mbox{\boldmath${\cal O}$} =
\mbox{\boldmath${\cal S}$},
\mbox{\boldmath${\cal P}$}, \\
&&\mu \frac{d}{d\mu} [\V_{jl}]
= - \sum_{k = 0}^{j} \lambda_{jk} [\V_{kl}] .
\end{eqnarray}
The fermionic operators $\V$ and $\U$ evolve with the same anomalous
dimensions, $\lambda_{jk}$, since $\U_{jl} = - \gamma_5 \V_{jl}$.
Note, that from symmetry properties of the operators it follows that
the anomalous dimensions for the parity even/odd operators vanish for
even/odd $j$, while they admit both values for fermionic case.

For the derivation of the Ward identities we proceed with the
regularization of the action via dimensional reduction since
otherwise $\delta^Q {\cal L}_{\rm cl} = - \frac{\g}{2} f^{abc} \bar\psi^a
\gamma_\mu \psi^b \bar\zeta \gamma_\mu \psi^c$ due to inapplicability
of the 4-dimensional Fiertz identity. Although it is a legitimate procedure
to use a symmetry breaking regulator, but this would result to an addendum
in the Ward identity and as a result in the relations we are going to
derive. With Siegel's method at hand the supersymmetry Ward identity
simply reads
\begin{equation}
\langle [ \mbox{\boldmath${\cal O}$}_{jl} ] \delta^Q \X \rangle
= -
\langle \delta^Q [ \mbox{\boldmath${\cal O}$}_{jl} ] \X \rangle ,
\end{equation}
with $\delta^Q \mbox{\boldmath${\cal O}$}_{jl}$ given by
Eqs.\ (\ref{SUSYtransfS})-(\ref{SUSYtransfV}). Finally, the dilatation
Ward identity \cite{BelMul98QCD} adjusted to the present scheme is
\begin{equation}
\langle [ \mbox{\boldmath${\cal O}$}_{jl} ] \delta^D \X \rangle
= \sum_{k = 0}^{j}
\left\{
( l + 3 ) \1 + \mbox{\boldmath${\gamma}$}^{\cal O} \right\}_{jk}
\langle [ \mbox{\boldmath${\cal O}$}_{kl} ] \X \rangle
+ \sum_i {\cal F}^i (\g)
\langle i [ \mbox{\boldmath${\cal O}$}_{jl} \Delta^i] \X \rangle ,
\end{equation}
where the last term stands for Green functions with the renormalized
operator insertions, i.e.\ differential vertex operator insertion
$[\Delta^\g] = \g \frac{\partial}{\partial\g} {\cal L}$ and equation 
of motion operators ${\mit\Omega}_\phi = \phi \frac{\delta S}{\delta
\phi}$. Here $l + 3 = l - 1 + 2 ( d^{\rm can}_G + 1 ) = l + 2
d^{\rm can}_Q$, for gluonic and fermionic operators, respectively,
with canonical dimension\footnote{Obviously, only ``good" components
are endowed with well defined canonical dimensions.} $d^{\rm can}_\phi$
of the field $\phi$. Thus from the last commutator\footnote{Note that
its nonzero r.h.s.\ just serves to shift the canonical scale dimension 
of the elementary field $\phi$ by $\frac{1}{2}$ in mass units, thus
changing, for instance $d^{\rm can}_G \to d^{\rm can}_G + \frac{1}{2}
= d^{\rm can}_Q$.} in Eq.\ (\ref{GradedAlgebra}) acting on the Green
function with conformal operator insertion (\ref{LCGreen}) $\langle
[ \mbox{\boldmath${\cal O}$}_{jl} ] [\delta^Q , \delta^D]_- \X \rangle
= \frac{1}{2} \langle [ \mbox{\boldmath${\cal O}$}_{jl} ] \delta^Q
\X \rangle$ and from the Ward equalities displayed above we can easily
obtain a set of identities for the anomalous dimensions. For this one
has to note that $\langle [ \mbox{\boldmath${\cal O}$}_{jl} \Delta^\g]
\X \rangle = \g \frac{\partial}{\partial\g} \langle [
\mbox{\boldmath${\cal O}$}_{jl}] \X \rangle$ and $\sum_{\phi = \psi, G}
\gamma_\phi \langle [ \mbox{\boldmath${\cal O}$}_{jl} {\mit\Omega}_\phi ]
\X \rangle = i \gamma_\phi N \langle [ \mbox{\boldmath${\cal O}$}_{jl} ]
\X \rangle$ \cite{BelMul98QCD} with $\gamma_\psi = \gamma_G = \gamma_\phi$
due to the gauge we have chosen. $N$ is the total number of fields
in $\X$. Therefore, these terms drop out from the commutator of the scale
and supersymmetry transformations and thus do not show up in the
constraints. Namely, for $\mbox{\boldmath${\cal O}$} =
\mbox{\boldmath${\cal S}$}$ we have in components
\begin{eqnarray}
\sum_{m = 0}^{n}
\left\{
{^{11}\!\gamma}^{\cal S}_{2n + 1, 2m + 1} [ \V_{2m, l} ]
+
{^{12}\!\gamma}^{\cal S}_{2n + 1, 2m + 1} [ \V_{2m + 1, l} ]
\right\}
\!\!\!\!&=&\!\!\!\!
\sum_{m = 0}^{n} \lambda_{2n, 2m} [ \V_{2m, l} ]
+
\sum_{m = 0}^{n - 1} \lambda_{2n, 2m + 1} [ \V_{2m + 1, l} ] \ ,
\nonumber\\
\sum_{m = 0}^{n}
\left\{
{^{21}\!\gamma}^{\cal S}_{2n + 1, 2m + 1} [ \V_{2m, l} ]
+
{^{22}\!\gamma}^{\cal S}_{2n + 1, 2m + 1} [ \V_{2m + 1, l} ]
\right\}
\!\!\!\!&=&\!\!\!\!
\sum_{m = 0}^{n} \lambda_{2n + 1, 2m} [ \V_{2m, l} ]
+
\sum_{m = 0}^{n} \lambda_{2n + 1, 2m + 1} [ \V_{2m + 1, l} ] \ ,
\nonumber
\end{eqnarray}
while for $\mbox{\boldmath${\cal O}$} = \mbox{\boldmath${\cal P}$}$
\begin{eqnarray}
\sum_{m = 0}^{n - 1}
{^{11}\!\gamma}^{\cal P}_{2n, 2m + 2} [ \U_{2m + 1, l} ]
+
\sum_{m = 0}^{n}
{^{12}\!\gamma}^{\cal P}_{2n, 2m} [ \U_{2m, l} ]
\!\!\!\!&=&\!\!\!\!
\sum_{m = 0}^{n - 1} \lambda_{2n - 1, 2m} [ \U_{2m, l} ]
+
\sum_{m = 0}^{n - 1} \lambda_{2n - 1, 2m + 1} [ \U_{2m + 1, l} ] \ ,
\nonumber\\
\sum_{m = 0}^{n - 1}
{^{21}\!\gamma}^{\cal P}_{2n, 2m + 2} [ \U_{2m + 1, l} ]
+
\sum_{m = 0}^{n}
{^{22}\!\gamma}^{\cal P}_{2n, 2m} [ \U_{2m, l} ]
\!\!\!\!&=&\!\!\!\!
\sum_{m = 0}^{n} \lambda_{2n, 2m} [ \U_{2m, l} ]
+
\sum_{m = 0}^{n - 1} \lambda_{2n, 2m + 1} [ \U_{2m + 1, l} ] \ .
\nonumber
\end{eqnarray}
Then, using linear independence of conformal operators we get
finally the following relations
\begin{eqnarray}
\label{S11}
{^{11}\!\gamma}^{\cal S}_{2n + 1, 2m + 1}
\!\!\!&=&\!\!\! {^{22}\!\gamma}^{\cal P}_{2n, 2m}
= \lambda_{2n, 2m} , \quad m \leq n , \\
\label{S12}
{^{12}\!\gamma}^{\cal S}_{2n + 1, 2m + 1}
\!\!\!&=&\!\!\! {^{21}\!\gamma}^{\cal P}_{2n, 2m + 1}
= \lambda_{2n, 2m + 1} , \quad  m \leq n - 1 , \\
\label{S21}
{^{21}\!\gamma}^{\cal S}_{2n + 1, 2m + 1}
\!\!\!&=&\!\!\! {^{12}\!\gamma}^{\cal P}_{2n + 2, 2m}
= \lambda_{2n + 1, 2m} , \quad m \leq n \ , \\
\label{S22}
{^{22}\!\gamma}^{\cal S}_{2n + 1, 2m + 1}
\!\!\!&=&\!\!\! {^{11}\!\gamma}^{\cal P}_{2n + 2, 2m + 2}
= \lambda_{2n + 1, 2m + 1} , \quad m \leq n \ , \\
\label{Dok}
{^{12}\!\gamma}^{\cal S}_{2n + 1, 2n + 1}
\!\!\!&=&\!\!\! 0 \ , \quad
{^{12}\!\gamma}^{\cal P}_{2n, 2n} = 0 \ .
\end{eqnarray}
This is the most general set of equations for the anomalous
dimensions of operators with total derivatives which arise in
a theory with ${\cal N} = 1$ supersymmetry.


\noindent {\it 5. Reduced supersymmetry relations.} From the above
relations for the anomalous dimensions of the operators which form
a supermultiplet we can deduce identities for the familiar anomalous
dimensions of quark and gluon operators. For this purpose note that
they are related to each other via the following matrix equation
\begin{equation}
\left(
\begin{array}{r}
\frac{1}{k}\,     {^{11}\!\gamma}_{jk} \\
\frac{1}{k + 1}\, {^{12}\!\gamma}_{jk} \\
\frac{1}{k}\,     {^{21}\!\gamma}_{jk} \\
\frac{1}{k + 1}\, {^{22}\!\gamma}_{jk}
\end{array}
\right)
= \frac{1}{2k + 3}
\left(
\begin{array}{cccc}
1                     & \frac{k + 3}{6}
& \frac{6}{j}         & \frac{k + 3}{j} \\
-1                    & \frac{k}{6}
& - \frac{6}{j}       & \frac{k}{j} \\
- \frac{j + 3}{j + 1} & - \frac{(k + 3)(j + 3)}{6(j + 1)}
& \frac{6}{j + 1}     & \frac{k + 3}{j + 1} \\
\frac{j + 3}{j + 1}   & - \frac{k(j + 3)}{6(j + 1)}
& - \frac{6}{j + 1}   & \frac{k}{j + 1}
\end{array}
\right)
\left(
\begin{array}{r}
{^{QQ}\!\gamma}_{jk} \\
{^{QG}\!\gamma}_{jk} \\
{^{GQ}\!\gamma}_{jk} \\
{^{GG}\!\gamma}_{jk}
\end{array}
\right) .
\end{equation}
Then the transformation to the usual operator basis becomes trivial.
Let us consider some particular limits of Eqs.\ (\ref{S11})-(\ref{S22}).
Namely, from Eqs.\ (\ref{Dok}) we obtain the well-known Dokshitzer
relations empirically established in the original paper \cite{Dok77}
for the vector channel (here and below $\gamma_{jj} \equiv \gamma_j$)
\begin{equation}
\label{DokSUSY}
{^{QQ}\!\gamma}^i_j + \frac{6}{j} {^{GQ}\!\gamma}^i_j
= \frac{j}{6} {^{QG}\!\gamma}^i_j + {^{GG}\!\gamma}^i_j ,
\quad i = V,A .
\end{equation}
From (\ref{S11}) and (\ref{S22}), reduced to the forward case, we have
\begin{equation}
\label{SUSY2}
{^{QQ}\!\gamma}^V_{j + 1}
+ \frac{6}{j + 1} {^{GQ}\!\gamma}^V_{j + 1}
=
{^{QQ}\!\gamma}^A_j
- \frac{j}{6} {^{QG}\!\gamma}^A_j
\quad\mbox{and}\quad
{^{QQ}\!\gamma}^A_{j + 1}
+ \frac{6}{j + 1} {^{GQ}\!\gamma}^A_{j + 1}
=
{^{QQ}\!\gamma}^V_j
- \frac{j}{6} {^{QG}\!\gamma}^V_j .
\end{equation}
Finally, from Eqs.\ (\ref{S12}) and (\ref{S21}) a relation follows
between the diagonal (read forward) and non-diagonal elements of
the anomalous dimensions of the conformal operators
\begin{eqnarray}
\label{for-nonfor}
&& \frac{6}{j} {^{GQ}\!\gamma}^V_j
- \frac{j + 3}{6} {^{QG}\!\gamma}^V_j
= \frac{j + 1}{2j + 1} \Delta^A_{j + 1, j - 1},
\quad
\frac{6}{j} {^{GQ}\!\gamma}^A_j
- \frac{j + 3}{6} {^{QG}\!\gamma}^A_j
= \frac{j + 1}{2j + 1} \Delta^V_{j + 1, j - 1}, \\
&&
\Delta^i_{j + 1, j - 1}
\equiv \frac{j - 1}{j + 1} {^{GG}\!\gamma}^i_{j + 1, j - 1}
+ \frac{j - 1}{6} {^{QG}\!\gamma}^i_{j + 1, j - 1}
- \frac{6}{j + 1} {^{GQ}\!\gamma}^i_{j + 1, j - 1}
- {^{QQ}\!\gamma}^i_{j + 1, j - 1} .
\end{eqnarray}
Obviously, in the leading order conformal operators do not mix
and the r.h.s.\ of the equations (\ref{for-nonfor}) are zero.
However, beyond one loop \cite{BelMul98QCD} these equations provide a
non-trivial check of existing results.


\noindent {\it 6. Transformation from SUSY preserving to conventional
DREG schemes.} Let us address now the question of explicit checks of
the above predictions beyond leading order of QCD
perturbation theory. Since all above equations hold only for
the entities evaluated by means of supersymmetry
preserving dimensional reduction we have to compute rotation matrices
to the conventional dimensional regularization\footnote{Remarkable that
the number of different transformation matrices in the forward case
equals to the number of independent investigations on the subject
\cite{Ell81}-\cite{Blu98}. Diversity of opinions is welcome but outside
physics.} (DREG) --- a scheme used in practical QCD calculations where
all higher order results are available. Note that the supersymmetric 
limit of ordinary QCD can be achieved by equating the Casimir operators
$C_A = C_F = 2 T_F = N_c$.

The change of the scheme is achieved via the following finite
transformation of the quark-gluon operator $\mbox{\boldmath${\cal O}$}
= \left( { {^Q\!{\cal O}} \atop {^G\!{\cal O}} } \right)$ renormalized
according to the conventional DREG to the DRED scheme
$[\mbox{\boldmath${\cal O}$}]^{\rm DRED} = \mbox{\boldmath$z$}
[\mbox{\boldmath${\cal O}$}]^{\rm DREG}$. Thus, the quark-gluon
anomalous dimension matrix, $\mbox{\boldmath$\gamma$}$, for the
regularization with DRED are related to the DREG one via
\begin{equation}
\mbox{\boldmath$\gamma$}^{\rm DRED}
= \mbox{\boldmath$z$}
\mbox{\boldmath$\gamma$}^{\rm DREG}
\mbox{\boldmath$z$}^{-1}
- \beta (\g) \frac{\partial}{\partial\g}
\mbox{\boldmath$z$} \cdot \mbox{\boldmath$z$}^{-1} .
\end{equation}

\begin{figure}[t]
\begin{center}
\vspace{4.2cm}
\hspace{-1.5cm}
\mbox{
\begin{picture}(0,220)(270,0)
\put(0,-30)                    {
\epsffile{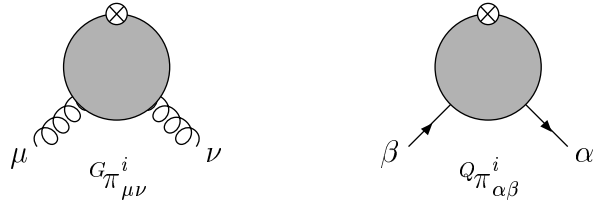}
                               }
\end{picture}
}
\end{center}
\vspace{-10.0cm}
\caption{\label{diagram} Generic form of Feynman diagrams for
$\mbox{\boldmath$z$}$ projected onto the tensor structures
given in the text.}
\end{figure}

In order to check the relations we have derived above in the two-loop
approximation the problem is thus reduced to the computation of
$\mbox{\boldmath$z$}$ at ${\cal O} (\alpha_s)$ (see Fig.\ \ref{diagram}).
Let us add few remarks on this calculation which has been performed with
ordinary QCD Feynman rules identifying afterwards the Casimir
operators. Since the Clifford algebra is considered as 4-dimensional
in the DRED as well as in conventional DREG the projectors used
are the same in both schemes ${^Q\!\pi}^{(V,A)} = \frac{1}{4} (1, \gamma_5)
\gamma_-$ for the parity even and odd sectors, respectively. On the
other hand the gluon polarization vectors are treated as 4-dimensional
in DRED and $d$-dimensional in DREG. We have then for parity even
case ${^G\!\pi}^{V}_{\mu\nu} = (d - 2)^{- 1} g^{\perp\, (d)}_{\mu\nu}$
with $d = 4$ for DRED and $d = 4 - 2 \epsilon$
for conventional DREG, while for odd parity ${^G\!\pi}^{A}_{\mu\nu} =
i (d - 2)^{- 1} (d - 3)^{- 1} \epsilon_{\mu\nu -+}$. We have used the 
HVBM scheme \cite{HVBM} for dealing with $\gamma_5$ and
$\epsilon_{\mu\nu\rho\sigma}$ which are pure 4-dimensional objects.
These procedure has proved to be the most reliable for these purpose.
Moreover, as a cross check we have adopted also Larin's prescription
\cite{Lar93} according to which we have used the substitution
$\gamma_\mu \gamma_5 = - \frac{i}{3!} \epsilon_{\mu\nu\rho\sigma}
\gamma_\nu \gamma_\rho \gamma_\sigma$ in quark-gluon diagrams and the
resulting product of two $\epsilon$-tensors has been understood as
$\epsilon_{\mu\nu\rho\sigma} \epsilon^{\alpha\beta\gamma\delta} =
- 4! g_{\mu}^{[\alpha} g_{\nu}^{\beta} g_{\rho}^{\gamma}
g_{\sigma}^{\delta]}$ with the metric tensors being $d$-dimensional for
DREG and 4-dimensional for DRED. Note, however, that we have been able
to obtain the same results as for the HVBM recipe only for $QG$ and $GG$
graphs and have failed for others. Due to gauge invariance of the
rotation matrices we have done the calculations using covariant
Feynman and non-covariant light-cone gauges with indeed identical
final results.

Finally, we have found the finite part of the anomalous dimensions of the
quark and gluon conformal operators, obtained from the difference of the
supersymmetry preserving and the conventional dimensional regularization
schemes, which reads
\begin{equation}
\mbox{\boldmath$z$}_{jk} = \1 \delta_{jk}
+ \frac{\alpha_s}{2 \pi} N_c
\left\{
\mbox{\boldmath$z$}^{\rm D}_j \delta_{jk}
+
\mbox{\boldmath$z$}^{\rm ND}_{jk}
\theta_{j - 2,k} [1 + (- 1)^{j - k}]
\right\} ,
\end{equation}
with the following diagonal matrices
\begin{equation}
\label{TransDiag}
\mbox{\boldmath$z$}^{{\rm D},V}_j
= \left(
\begin{array}{cc}
- \frac{j(j + 3)}{2(j + 1)(j + 2)}
&
\frac{12}{j(j + 2)(j + 3)}
\\
\frac{j}{6 (j + 2)}
&
-\frac{1}{6}
\end{array}
\right) , \quad
\mbox{\boldmath$z$}^{{\rm D},A}_j
= \left(
\begin{array}{cc}
- \frac{j(j + 3)}{2(j + 1)(j + 2)}
&
\frac{12}{j(j + 1)(j + 2)}
\\
- \frac{j}{3 (j + 1)(j + 2)}
&
-\frac{1}{6} - \frac{4}{(j + 1)(j + 2)}
\end{array}
\right) ,
\end{equation}
for the vector and axial channels, respectively, and the universal
non-diagonal part
\begin{equation}
\label{TransNonDiag}
\mbox{\boldmath$z$}^{\rm ND}_{jk}
= \left(
\begin{array}{cc}
0
&
\frac{6 (2k + 3)}{k( k + 1 )( k + 2 )( k + 3 )} \\
- \frac{(2k + 3)}{6( k + 1 )( k + 2 )}
&
- \frac{(2k + 3)(j - k)(j + k + 3)}{k(k + 1)(k + 2)(k + 3)}
\end{array}
\right) .
\end{equation}

In the course of the calculation we have clarified the reason for a
difference in the rotation matrices given in the literature for forward
scattering. For instance, the recent results of Ref.\ \cite{Blu98} can be
reproduced in practically all cases (except axial $GQ$ channel) provided
we discard the contributions of $\epsilon$-scalars in the external
lines. We can hardly advocate this recipe since the
$\mbox{\boldmath$z$}$-matrices have to be considered as insertions into
the internal virtual lines, so that the $\epsilon$-scalars contribute on
equal footing with the $d$-dimensional gauge particles. The diagonal
eigenvalues, $\mbox{\boldmath$z$}^{\rm D}_j$, coincide with
\cite{Ell81,Kun94} for parity even and with \cite{Vog96} for parity odd
operators.

Note that only diagonal elements of these matrices are required
to fulfill the supersymmetry relations in Eqs.\ (\ref{DokSUSY})
and (\ref{SUSY2}) which can be explicitly checked transforming the
results of Refs.\ \cite{MerNee96,Vog96,FurPet80} with the help of
the rotation matrices (\ref{TransDiag}) to the DRED scheme. In Eq.\ 
(\ref{for-nonfor}) the off-diagonal elements \cite{BelMul98QCD}
enter which have to be rotated with (\ref{TransNonDiag}) to the
supersymmetry preserving scheme as well. The appearance of these 
turns out to be the reason why the authors of Ref.\ \cite{Blu98} have
found the violation of the third SUSY relation in NLO --- they have 
used an equation of \cite{BFKL85} which corresponds to our Eq.\ 
(\ref{for-nonfor}) but with zero r.h.s.

Moreover, we have found that the general relations (\ref{S11}-\ref{S22})
are satisfied when transformed with the rotation matrices we have derived
presently. This provides a confirmation for the correctness on the 
two-loop anomalous dimensions for the QCD composite operators available 
in the literature \cite{BelMul98QCD,MerNee96,Vog96,FurPet80}.


\noindent {\it 7. Summary.} To summarize we have derived in the present
study a general set of relations for the anomalous dimensions of
the bosonic and fermionic conformal composite operators in the
Yang-Mills theory with ${\cal N} = 1$ supersymmetry. In their reduced
form two of them were known earlier while the last one
(Eq.\ (\ref{for-nonfor})) contains a novel non-vanishing r.h.s.\ with the
non-diagonal elements. This explains the difficulties in the check of 
their validity beyond leading order of perturbation theory observed 
before. We have thus supported our results for the non-diagonal NLO 
anomalous dimensions of the conformal operators derived in 
\cite{BelMul98QCD} and we now have a stronger evidence for the 
supersymmetric nature of the universality of the special conformal 
anomalies of conformal operators conjectured there. This issue is 
currently under study \cite{BelMulprep}. 

\vspace{0.5cm}

This work was supported by BMBF and the Alexander von Humboldt
Foundation (A.B.).

\vspace{1cm}


\noindent {\it Appendix.} Here we briefly describe the main steps for
the construction of an irreducible representation of supersymmetry in
the basis of the conformal composite operators. The variation of the
bosonic conformal operators in Eq.\ (\ref{treeCO}) leads to
\begin{eqnarray}
\delta^Q\,
\left\{\!\!\!
\begin{array}{c}
{^Q\!{\cal O}}^V \\
{^Q\!{\cal O}}^A
\end{array}
\!\!\!\right\}_{jl}
\!\!\!\!\!&=&\!\!\! \left[ 1 - (-1)^j
\left\{
{ \phantom{-}1 \atop -1 }
\right\} \right]
\frac{(-1)(j + 2)}{2} \
\bar\zeta G^{a \perp}_{+ \mu} (i \partial_+)^l
P^{(1,1)}_j \left(
\frac{\stackrel{\leftrightarrow}{\D}_+}{\partial_+}
\right)
\gamma^\perp_\mu
\left\{\!\!\!
\begin{array}{c}
1 \\
\gamma_5
\end{array}
\!\!\!\right\}
\psi^a_+, \nonumber\\
\delta^Q\,
\left\{\!\!\!
\begin{array}{c}
{^G\!{\cal O}}^V \\
{^G\!{\cal O}}^A
\end{array}
\!\!\!\right\}_{jl}
\!\!\!\!\!&=&\!\!\! \left[ 1 - (-1)^j
\left\{
{ \phantom{-}1 \atop -1 }
\right\} \right]
\frac{(j + 2) (j + 3)}{12} \
\bar\zeta G^{a \perp}_{+ \mu} (i \partial_+)^{l - 1}
i\!\stackrel{\rightarrow}{\D}_+
P^{(2,2)}_{j-1} \left(
\frac{\stackrel{\leftrightarrow}{\D}_+}{\partial_+}
\right)
\gamma^\perp_\mu
\left\{\!\!\!
\begin{array}{c}
1 \\
\gamma_5
\end{array}
\!\!\!\right\}
\psi^a_+ . \nonumber
\end{eqnarray}
On the other hand since the fermionic operator, $(G\psi)_{jj}$, like
the bosonic ones (\ref{treeCO}) has to be the highest weight vector of the
corresponding conformal tower, $[(G\psi)_{jj}, \K_-]_- = 0$, with the
generator of the special conformal transformation ${\cal K}_\mu$, this
requirement fixes the indices of the Jacobi polynomial to be $d_\phi +
s_\phi - 1$. Here $d_\phi$ and $s_\phi$ are the canonical scale dimension
and spin of the field $\phi$, respectively. For the case at hand we have
thus $P^{(2,1)}_j$. This condition implies severe constraints on the
coefficients with which the operators (\ref{treeCO}) can enter ${^Q\!C_j} 
{^Q\!{\cal O}}_{jl} + {^G\!C_j} {^G\!{\cal O}}_{jl}$, which form an 
irreducible representation of the supersymmetry algebra. The corresponding
equation for ${^i\!C_j}$ has two solutions. They are given in Eq.\
(\ref{SUSYmultiplet}).

\end{document}